\documentstyle[psfig,twoside]{article}
\def \ref#1{{\par\noindent \hangindent=3em \hangafter=1
     \advance \rightskip by 3em #1.\par}}

\begin{document}
\begin{Large}
{\bf Electromagnetic radiation and motion of arbitrarily shaped particle} \\
\end{Large}

{\sl J. Kla\v{c}ka} \\\\
Astronomical Institute, Faculty of Mathematics, Physics, and Informatics,
Comenius University, \\ 
Mlynsk\'{a} dolina, 842 48 Bratislava, Slovak Republic\\

\vspace{5mm}
{\bf Abstract:}
Covariant form of equation of motion for arbitrarily shaped particle
in the electromagnetic radiation field is presented. 
Equation of motion in the proper frame of the particle uses the
radiation pressure cross section 3 $\times$ 3 matrix. 
The obtained equation of motion is compared with known result.
 
\vspace{5mm}
{\bf 1. Introduction}

Dealing with motion of real dust particles in the Universe, it may be
important to consider action of electromagnetic radiation. Usage of correct
equation of motion is inevitable, in that case. 

Poynting-Robertson effect (Robertson 1937) was used for several decades as a 
model equation of motion. However, since 1994 (Kla\v{c}ka 1994) equations
of motion for real, arbitrarily shaped, particle were presented. The most
complete form was presented in Kla\v{c}ka (2000a) -- covariant form of equation
of motion. Later on, approximations to the first order in $v/c$
($v$ -- velocity of the particle with respect to the source of radiation,
$c$ -- speed of light) for dust particles were presented: Kla\v{c}ka (2000b),
Kla\v{c}ka and Kocifaj (2001). Application to larger bodies may be found in
Kla\v{c}ka (2000c).

This paper presents covariant equation of motion of the particle in the
electromagnetic radiation field for a little different initial formulation.
It is supposed that interaction between the particle and electromagnetic 
radiation is expressed in term of radiation pressure cross section matrix.

\vspace{5mm}
{\bf 2. Formulation of the problem}

Let the equation of motion of a particle in the electromagnetic radiation
field is expressed in the form
\begin{eqnarray}\label{1}
\frac{d \vec{p'}}{d \tau} &=& \frac{1}{c} ~S'~\left ( C'~\hat{\vec{S'_{i}}}
                           \right )
\nonumber \\
\frac{d E'}{d \tau} &=& 0 ~, 
\end{eqnarray}
in the proper frame of the particle; $S'$ is the flux density of the radiation
energy, $C'$ is pressure cross section 3 $\times$ 3 matrix, 
$\hat{\vec{S'_{i}}}$ is unit vector of the incident radiation, $\tau$ is
proper time, $c$ is the speed of light.

We are interested in deriving equation of motion of the particle in the 
rest frame of the source: the particle moves with instantaneous velocity
$\vec{v}$ with respect to the source, the unit vector of the incident
radiation is $\hat{\vec{S_{i}}}$ and other physical quantities measured
in the rest frame of the source are also unprimed. 

\vspace{5mm}
{\bf 3. Reformulation of the initial equation of motion}

Let the components of the pressure cross section $C'$ 3 $\times$ 3 matrix
are given in a basis of orthonormal vectors
$\hat{\vec{e'_{b1}}}$, $\hat{\vec{e'_{b2}}}$, $\hat{\vec{e'_{b3}}}$.

Let us define a new orthonormal basis of vectors
$\hat{\vec{S'_{i}}}$, $\hat{\vec{e'_{1}}}$, $\hat{\vec{e'_{2}}}$. We may
write  
\begin{equation}\label{2}
\hat{\vec{S'_{i}}} = \sum_{k=1}^{3} ~ s'_{k} \hat{\vec{e'_{bk}}} ~,~~
\hat{\vec{e'_{1}}} = \sum_{k=1}^{3} ~ p'_{k} \hat{\vec{e'_{bk}}} ~, ~~
\hat{\vec{e'_{2}}} = \sum_{k=1}^{3} ~ q'_{k} \hat{\vec{e'_{bk}}} ~.
\end{equation}

Using Eqs. (2) we have
\begin{equation}\label{3}
C'~\hat{\vec{S'_{i}}} = \sum_{k=1}^{3} ~ \sum_{l=1}^{3} ~ C'_{kl} ~ s'_{l}
                        \hat{\vec{e'_{bk}}} ~. 
\end{equation}

On the basis of Eqs. (2) we can write
\begin{equation}\label{4}
\hat{\vec{e'_{bk}}} =  s'_{k} \hat{\vec{S'_{i}}} ~+~ 
                       p'_{k} \hat{\vec{e'_{1}}} ~+~
                       q'_{k} \hat{\vec{e'_{2}}} ~, ~~~k = 1, 2, 3  ~.
\end{equation}

Putting Eqs. (4) into Eq. (3) we obtain 
\begin{eqnarray}\label{5}
C'~\hat{\vec{S'_{i}}} &=& \sum_{k=1}^{3} ~ \sum_{l=1}^{3} ~ C'_{kl} ~ s'_{l} ~
                        \left (  s'_{k} \hat{\vec{S'_{i}}} ~+~
                                 p'_{k} \hat{\vec{e'_{1}}} ~+~
                                 q'_{k} \hat{\vec{e'_{2}}} ~ \right ) =
\nonumber \\
&=& \left ( \sum_{k=1}^{3} ~ \sum_{l=1}^{3} ~ 
            C'_{kl} ~ s'_{l} ~s'_{k} \right ) ~ \hat{\vec{S'_{i}}} ~+~                
    \left ( \sum_{k=1}^{3} ~ \sum_{l=1}^{3} ~ 
            C'_{kl} ~ s'_{l} ~p'_{k} \right ) ~ \hat{\vec{e'_{1}}} ~+~
    \left ( \sum_{k=1}^{3} ~ \sum_{l=1}^{3} ~ 
            C'_{kl} ~ s'_{l} ~q'_{k} \right ) ~ \hat{\vec{e'_{2}}} ~.                              
\end{eqnarray}

On the basis of Eq. (5) we can rewrite the first of Eqs. (1) to the form
\begin{equation}\label{6}
\frac{d \vec{p'}}{d \tau} = \frac{1}{c} ~S'~\left \{
\left ( \hat{\vec{S^{'T}_{i}}} ~C'~\hat{\vec{S'_{i}}} \right )
~\hat{\vec{S'_{i}}} ~+~
\left ( \hat{\vec{e^{'T}_{1}}} ~C'~\hat{\vec{S'_{i}}} \right )
~\hat{\vec{e'_{1}}} ~+
\left ( \hat{\vec{e^{'T}_{2}}} ~C'~\hat{\vec{S'_{i}}} \right )
~\hat{\vec{e'_{2}}} ~  \right \}  ~.
\end{equation}

\vspace{5mm}
{\bf 4. Covariant form of equation of motion}

Comparison of our Eq. (6) with Eq. (7) in Kla\v{c}ka (2000a) enables
immediately write covariant form of equation of motion, on the basis
of Eq. (28) in Kla\v{c}ka (2000a): 
\begin{equation}\label{7}
\frac{d \vec{p^{\mu}}}{d \tau} = \frac{w^{2} ~S}{c^{2}} ~\left \{
\left ( \hat{\vec{S^{'T}_{i}}} ~C'~\hat{\vec{S'_{i}}} \right )~
\left ( c ~b_{i}^{\mu} ~-~ u^{\mu} \right ) ~+~ \sum_{j=1}^{2} ~ 
\left ( \hat{\vec{e^{'T}_{j}}} ~C'~\hat{\vec{S'_{i}}} \right )~
\left ( c ~b_{j}^{\mu} ~-~ u^{\mu} \right ) ~  \right \}  ~,
\end{equation}
where $p^{\mu} = m~u^{\mu}$, $u^{\mu} = (\gamma ~c, ~\gamma ~\vec{v} )$,
$w = \gamma ~( 1 ~-~\vec{v} \cdot \hat{\vec{S_{i}}} ~/ ~c )$,
$b_{i}^{\mu} = ( 1~/ ~w, \hat{\vec{S_{i}}}~/ ~w )$,
$b_{j}^{\mu} = ( 1~/ ~w_{j}, \hat{\vec{e_{j}}}~/ ~w_{j} )$,
$w_{j} = \gamma ~( 1 ~-~\vec{v} \cdot \hat{\vec{e_{j}}} ~/ ~c )$.


\vspace{5mm}
{\bf 5. Conclusion}

We have obtained covariant equation of motion for the particle
moving in electromagnetic radiation field. 

We have supposed that interaction between the particle and electromagnetic 
radiation is described by radiation pressure cross section $C'$ 3 $\times$ 3 
matrix. Simple transformations reduce this mathematical form of description
to that presented in Kla\v{c}ka (2000a). This has enabled to write
final Eq. (7).

{\bf Acknowledgements:} This work was supported by the VEGA grant
No.1/7067/20. 

\vspace{3mm}
\begin{large}
{\bf References}
\end{large}

\ref{Kla\v{c}ka J., 1994, Earth, Moon, and Planets 64, 125-132} 
\ref{Kla\v{c}ka J., 2000a, http://xxx.lanl.gov/abs/astro-ph/0008510}
\ref{Kla\v{c}ka J., 2000b, http://xxx.lanl.gov/abs/astro-ph/0009108}
\ref{Kla\v{c}ka J., 2000c, http://xxx.lanl.gov/abs/astro-ph/0009109}
\ref{Kla\v{c}ka J., Kocifaj M., 2001, JQSRT 70/4-6, 595-610} 
\ref{Robertson, H. P., 1937, MNRAS 97, 423-438} 

\end{document}